\documentclass[twocolumn]{svjour3}          
\usepackage{graphicx}
\usepackage{adjustbox}
\usepackage{amsmath}
\usepackage{epstopdf}
\usepackage[graphicx]{realboxes}
\usepackage{rotating}
\usepackage{float}
\usepackage{latexsym}
\usepackage{amssymb}
\usepackage{array}
\usepackage{longtable}
\usepackage{dcolumn}
\usepackage{bm}

\begin{document}

\title{Emergent Universe Scenario in Modified Gauss-Bonnet Gravity }

\author{B. C. Paul  \and 
              S. D. Maharaj  \and
             A. Beesham }

\institute{B. C. Paul \\ 
              \email{bcpaul@associates.iucaa.in}\at
                  Department of Physics, University of North Bengal, Siliguri, Dist. : Darjeeling 734 014, West Bengal, India \\
S. D. Maharaj \\
           \email{maharaj@ukzn.ac.za}\at
           Astrophysics and Cosmology Research Unit, School of Mathematics, Statistics and
Computer Science, University of KwaZulu-Natal, Private Bag X54001, Durban 4000, South
Africa \\
           A. Beesham \\
             \email{abeesham@yahoo.com}\at
              Department of Mathematical Sciences, University of Zululand \\
Private Bag X1001, Kwa-Dlangezwa 3886, South Africa \\
             }
             
\date{Received: date / Accepted: date}	
\maketitle							

\vspace{0.5in}

\begin{abstract}
We present  modified Gauss-Bonnet gravity without matter in four dimensions which accommodates flat emergent universe (EU) obtained in Einstein's general theory of gravity  with a non-linear equation of state.  The  EU model is interesting which is free from big-bang singularity with other observed features of the universe. It is assumed that the present universe  emerged out from a  static Einstein universe phase exists in the infinite past.  To obtain a  flat EU model we reconstructed mimetic modified $f(G)$-gravity ($G$ representing Gauss-Bonnet terms) without matter.  The functional form of $f(G)$-gravity is determined  which accommodates the early inflation and late accelerating phases without matter. 

\end{abstract}



\section{. \; Introduction}

In modern cosmology it is believed that the present universe  emerged from an inflationary phase in the early universe \cite{kolb,kolb1,kolb2}. The observational cosmology in recent decades  predicted  that our universe emerged out from an early inflationary epoch. Inspite of success of  hot Big Bang model of the universe it fails to find no solution in Einstein's general theory of relativity with perfect fluid. 
The concept of early inflation  introduced in cosmology resolve some of these  issues in cosmology. It is known that inflation can be obtained in  a semiclassical theory of gravity or in a  higher order  theory of gravity.
Recent cosmological observations~\cite{riess,riess1,riess2} predicted   another interesting feature of the 
 universe  that the  present universe is passing through an accelerating phase.
Theoretically it is challenging to describe the origin of  the late accelerating phase as the physics of the inflation~\cite{guth,guth1,guth2,guth3,guth4} and the introduction of 
a small cosmological constant for late acceleration are  among the features that are not yet 
understood~\cite{sean,sean1}. A number of proposal put forwarded in understanding the late accelerating universe similar in  this direction to  realize the early inflation either by  a modification of the gravitational sector ~\cite{ca20,suj,suj1}  or by a modification of the  matter sector of the Einstein's general theory of relativity. Thus  there is enough motivation  to explore for  an alternative cosmological model. 
Astashenok {\it et al.} \cite{asc} in a mimetic modified
gravity models investigated various scenarios of cosmological evolution, with or
without extra matter fluids to obtain different cosmological evolutions. Considering a simplest formulation based on the use of the
Lagrange multiplier constraint it is shown that in some  theories, it is possible
to realise accelerated expansion of the Universe or even unified evolution,
which includes inflation with dark energy, and dark matter described by the theory. 
Adopting  general reconstruction schemes, it is found that mimetic gravity also generates a specific
cosmological evolution namely,   bouncing cosmologies.
Recently, in the literature \cite{asc,bam9,kb} cosmological models with a number of known evolutionary features are examined. The motivation of the paper is to reconstruct  modified Gauss-Bonnet gravity which permits a singularity free cosmology namely, flat emergent universe model which encompass the unified evolutions of the universe.

  The possibility of  a cosmological model~\cite{har} in 
which there is no Big Bang singularity,  no beginning of time, and the universe  
effectively avoids a quantum regime for space-time by staying large at all 
times was considered first by Ellis and Maartens~\cite{ellis} satisfactorily.  In  the model the universe started out in the infinite past in an almost static Einstein universe  and subsequently  it entered in an  expanding phase slowly, 
eventually evolving into a hot Big Bang era. Later Ellis {\it et al.} ~\cite{e1} constructed an emergent universe scenario for a spatially closed universe with a minimally coupled scalar field $\phi$, using a special 
form for the interaction potential  $V(\phi)$. However, it was shown \cite{ellis2} immediately that  such a
 potential can be recovered starting from a 
modified gravitational action with a polynomial Lagrangian $L= R + \alpha R^2$ 
making use of a suitable conformal transformation and identification of the field as $\phi= - \sqrt{3} \; \ln (1+ 2 \alpha R) $.  In the above however  one requires to have  a negative coupling term $\alpha$ for a consistent emergent universe (EU)  scenario. Later in a flat space-time Mukherjee {\it et. al.} \cite{sba} obtained a new cosmological solutions with a static Einstein universe in the framework of a semiclassical gravity \cite{staro}. 

It is known that EU scenario 
 promises to solve several conceptual and technical 
issues of the Big Bang model.  Recent cosmological observations predict that the 
universe is most likely to be spatially flat.   Therefore, it is important to explore a flat EU  model.
 A decade ago Mukherjee {\it et al.}~\cite{euorg}  in the framework of Einstein gravity  obtained an EU scenario in a flat universe   making use of  cosmic fluid that describe by a non-linear equation of state $
 p = A \rho- B \sqrt{\rho} $ 
where $A$ and $B$ are arbitrary constants. 

The possibility of emergent cosmology  using the effective potential formalism has been further explored recently showing that new models of emergent 
cosmology  can be realized within a limit set by the constraints imposed by the cosmic microwave background (CMB) radiation ~\cite{bag}. It has also
been  demonstrated that within the framework of
modified gravity, the emergent scenario may be obtained in a spatially open or closed universe.
It is shown that  a static Einstein universe   solution permitted in a modified theories of gravity with  perfect fluid~\cite{e1} may be  useful for constructing an alternative 
cosmological model without Big Bang singularity.
It is also shown later that   such an EU scenario can be 
realized  even in a flat universe  in a modified theory of gravity by including higher order
Gauss-Bonnet terms in the presence of a dilaton coupling~\cite{eugb}. Subsequently the EU model in a flat universe is explored in various
theoretical framework, namely Brane 
world ~\cite{b1,b2,b3}, Brans-Dicke theory~\cite{campo}, as well as in the context of the
non-linear sigma model~\cite{bee} and found that it can be obtained with all the features of the observed universe. The EU model accommodates late time 
de-Sitter expansion and thus, naturally permits  late time acceleration of
the universe. Such a scenario is promising from the perspective of
offering unified early as well as late time dynamics of the universe. Note
however, that the focal point of unification in such emergent universe models
lies in the choice of  equation of state for the polytropic fluid, while 
several other models of unification rely more on the scalar field dynamics
in the presence of a suitable field potential~\cite{asm1,asm2,asm3}. A number of issues pertaining to
different models of EU have  been discussed in the literature~\cite{cim,bag}. It is known that Gauss-Bonnet term without dilaton field does not contribute in the dynamics \cite{eugb} of the universe in four dimensions as it is a Euler number. The motivation  of the paper is to obtain EU scenario in a modified gravity with a modified Gauss-Bonnet gravity term without dilaton field which can accommodate  a flat EU model of the universe {\it i.e.}, we reconstruct  theoretically an acceptable modified gravity with GB terms in the gravitational action for accommodating EU scenario.  The Emergent Universe model is interesting because of its various observed features. It permits a non-singular solution of cosmology,  an important and new property  in understanding the early evolution of the universe, which is different from that obtained recently in the literature \cite{asc,asc1} where a singular bounce solution is obtained in a $F(G)$ theory of gravity where $G= R^2-4 R_{\mu \delta} R^{\mu \delta} + R_{\mu \alpha \nu \beta}R^{\mu \alpha \nu \beta}$  is the Gauss-Bonnet combination.

The paper is organized as follows: In Sec. 2, we set up the relevant field equation  starting from a gravitational action for  modified Gauss-Bonnet gravity. In Sec. 3, we present the Gauss-Bonnet modified theory for emergent universe,  
In Sec. 4, non-Singular bounce from mimetic    $F(G)$ gravity is presented and in Sec. 5, we present a brief discussion.

\section{.  \; Gravitational Action and the Field Equations}

The gravitational action for modified Gauss-Bonnet  gravity theory   \cite{bam5,bam6,bam7,bam8,sj,sj1,sj2,sj3,bam4} is given by 
\begin{equation}
\label{e1}
I = - \frac{1}{2 \kappa^2} \int \left( R + F(G)  \right) \sqrt{- g} \; \;  d^4x  + I_m
\end{equation}
where $\kappa^2=  \frac{1}{M_{P}^2}$ with the Planck mass represented by $M_P$, $ R$ represents the Ricci scalar and $I_m$ represents the  action for matter fields. The field equation is obtained by varying the metric which is given by 
\[
R_{\mu \nu} - \frac{1}{2} g_{\mu \nu} R - \frac{1}{2}  g_{\mu \nu} F(G) - (- 2 R R_{\mu \nu} + 4 R_{\mu \delta} R_{\nu}^{\delta} - 2 R_{\mu}^{\delta \sigma \tau} R_{\nu \delta \sigma \tau} 
\]
\[
\hspace{0.8 cm} 
+ 4 g^{\alpha \delta} g^{\beta \sigma} R_{\mu \alpha \nu \beta} R_{\delta \sigma} ) F'(G)  - 2 (\triangledown_{\mu} \triangledown{\nu} F'(G) ) R 
\]
\[
\hspace{0.6 cm}  
+ 2 g_{\mu \nu} (\square F'(G)) R - 4 (\square F'(G)) R_{\mu \nu} +4(\triangledown_{\mu} \triangledown_{\nu} F'(G) ) R^{\delta}_{\nu} 
 \]
 \[
 + 4 (\triangledown_{\delta} \triangledown_{\nu} F'(G)) R^{\delta}_{\mu} -4 g_{\mu \nu} (\triangledown_{\delta} \nabla_{\sigma} F'(G)) R^{\delta  \sigma} 
\]
\begin{equation}
\label{e2}
+ 4 (\triangledown_{\delta} \triangledown_{\sigma} F'(G)) g^{\alpha \delta} g^{\beta \sigma} R_{\mu \alpha \nu \beta} = \kappa^2 T_{\mu \nu}
\end{equation}
where $T_{mu \nu}$ is the energy momentum tensor.

We consider a  Robertson-Walker metric for a spatially homogeneous and isotropic space-time which is given by 
\begin{equation}
\label{e3}
ds^2= -dt^2+ a^2(t)\left[ \frac{dr^2}{1- k r^2}+ r^2 \left( d\theta^2 + sin^2 \theta \;  d\phi^2 \right) \right]
\end{equation} 
in the unit $c=1$ and $\hbar=1$, where  $a (t)$ represents the scale factor of the universe and $k =0, + 1, -1 $ represent flat, close and open universe respectively. As we explore emergent universe in a flat space-time \cite{euorg}
 we consider here $k=0$. 
The Ricci scalar and the Gauss-Bonnet invariant in a flat space-time are given by
\begin{equation}
\label{e4}
R= 6 \dot{H} + 12 H^2, \;\;\;\;\; G= 24 H^2 (\dot{H} + H^2)
\end{equation}
where $H=\frac{\dot{a}}{a}$ is the Hubble parameter and $\dot{()}$ is the time derivative.
The gravitational field equations are given by
\begin{equation}
\label{e5}
6 H^2 + F(G) - G F'(G) + 24 H^3 \dot{G} F'' (G) =  2 \kappa^2 \rho
\end{equation}
\[
4\dot{H} + 6 H^2 +F(G) - GF'(G) + 16H \dot{G} (\dot{H} +H^2) f''(G)
\]
\begin{equation}
\label{e6}
\hspace{1.0 cm}    + 8 H^2 \ddot{G} F''(G) + 8 H^2 \dot{G}^2 F'''(G) = -2 \kappa^2 p
\end{equation}
In four dimensions emergent universe (EU) can be obtained with a non-linear equation of state in Einstein gravity. The EU scenario in a flat universe was obtained by 
Mukherjee {\it et al.} \cite{euorg} which evolves out of a static phase of the universe in the 
infinitely past time. It  is free from big bang singularity and exhibits the features of an observable universe \cite{pt1,pt2,bc2,pt3,bas,ab}.

It has been shown that the EU model can be implemented in a modified gravity with Gauss-Bonnet terms coupled with dilaton field \cite{eugb}. In this paper we explore EU model in vacuum in a modified Gauss-Bonnet gravity and determine the functional form of $f(G)$ in the action. It is known that in 4 dimensions only GB terms in the action is a Euler number which is not contributing in the dynamics. However, it is shown that modified Gauss-Bonnet gravity plays an important role in understanding the evolution of the universe  (\cite{bam}-\cite{bam9}). The form of  $F(G)$-gravity will be reconstructed by using the method given in Ref. \cite{gc,gc1,kb} in the next section.

\subsection{ \bf Reconstruction Method of $F(G)$ gravity}

We reconstruct $F(G)$ gravity models by using the method of Ref. (\cite{bam8,gc,gc1}). 
Consider the gravitational action  \cite{bam9} without matter
\begin{equation}
\label{e7}
I = - \frac{1}{2 \kappa^2} \int \left( R +  P(t) G + Q(t) \right) \sqrt{- g} \; \;  d^4x 
\end{equation}
where $P(t)$ and $Q(t)$ are two proper functions of a scalar field $t$, which is interpreted as the cosmic time. Now varying the action with respect to $t$, we obtain  \cite{bam8,gc,gc1,kb,ca,ca1}
\begin{equation}
\label{e8}
\frac{dP(t)}{dt} G + \frac{dQ(t)}{dt} =0.
\end{equation}
The solution of the differential equation  can be expressed as $t=t(G)$. Consequently substitution of  the above solution in eq.  (\ref{e7}), yields the modified Gauss-Bonnet gravity functional in the Einstein-Hilbert action
\begin{equation}
 \label{e9}
 F(G) = P(t) G + Q(t).
 \end{equation}
It is interesting to determine $P(t)$ and $Q(t)$ in terms of Hubble parameter to study  observed cosmological consequences.   The modified action  may work well to get early inflation as well as late accelerating phase.
 
Using eqs. (\ref{e5}) and (\ref{e9}) one obtains
\begin{equation}
 \label{e10}
 Q(t) = - 6 H^2(t) - 24 H^3 (t) \frac{dP(t)}{dt}.
 \end{equation}
Finally using eqs. (\ref{e6}), (\ref{e9}) and (\ref{e10}), one obtains a second order differential equation given by
\begin{equation}
\label{e11}
2 H^2 (t) \frac{d^2 P(t)}{dt^2}  + 2 H(t) \left( 2 \dot{H}(t) - H^2(t) \right) \frac{dP(t)}{dt}  + \dot{H}=0 .
\end{equation}

\section{. \; Gauss-Bonnet Modified Gravity and The Emergent Universe}

The energy-density and pressure obtained from the Einstein field equations in 4-dimensions are  given by 
 \begin{equation}
 \label{e12}
 \rho = 3 H^2,  \; \; \; \; p = - (2 \dot{H}+2 H^2)
 \end{equation}
 where $H$ represents Hubble parameter where $\kappa^2=1$. The field equations with a non-linear   equation of state (EoS) :
  $p= A \rho - B \sqrt{\rho}$  was considered by Mukherjee {\it et. al.} \cite{euorg}  to obtain 
an emergent universe (EU)  model. The important feature of the EU model is that there is no-singularity and it describes the observed features of the universe. The conservation equation  is given by
	\begin{equation}
	 \label{e13}
	\frac{d\rho}{dt} + 3 H (\rho + p) = 0.
	\end{equation}
	The conservation equation  can be integrated using the non-linear EoS  which determines the energy density in terms of the scale factors. The   energy density is 
	\begin{equation}
	\label{e14}
	\rho(a) = \frac{1}{(A+1)^{2}}\left(B+ \frac{K}{a^{\frac{3(A+1)}{2}}}\right)^{2}
	\end{equation}
	which is new and interesting it depends on the EoS parameters  $A$ and $B$ and $K$ is an integration constant. It may be mentioned here that Chaplygin gas  \cite{chap} and its modified forms \cite{chap1,chap2,chap3} given by $p= A \rho - \frac{B}{\rho^{\alpha}}$ with $\alpha > 0$ which is also a promising candidate in cosmology does not permit such an expression for energy density which can be identified with different types of cosmic fluids. In the non-linear EoS considered by us it permits three different types of  fluids   as the composition of cosmic fluids determined by the parameter $A$.
	Expanding the above equations,  the total energy density can be identified with  three different fluids  corresponding to three different types of EoS. Therefore, one can express
	\begin{equation}
	\rho(a) = \Sigma_{i=1}^{3} \rho_{i} \;\;and\;\; p(a) = \Sigma_{i=1}^{3} p_{i}.
	\end{equation}
	It is noted that the parameter $A$ plays an important role in understanding different types of cosmic fluids present in the universe  \cite{euorg}. Later considering interactions among the fluids it is shown that a physically viable emergent universe  can be obtained for different strengths of interactions among them accommodating observed universe \cite{bas}. 
	The scale factor of the universe is obtained from eq. (\ref{e12}) which is
\begin{equation}
\label{e42}
	a(t) = \left[ \frac{3 K (A+1)}{2} \left( \sigma + \frac{2}{\sqrt{3} B} e^{\frac{\sqrt{3}}{2} }B t \right) \right ] ^{\frac{2} {3(A+1)}}
\end{equation}
where $\sigma$ and $K$ are integration constants.	 
	It is evident that at  infinitely past {\it i.e.} $t \rightarrow - \infty$, the scale factor becomes $a= a_o$ a static Einstein universe which evolves later to a dynamical universe scenario \cite{euorg,bas} encompassing the present observed universe.
	The interesting aspect of the emergent universe is that there is no singularity, early inflationary universe can be realized in addition to present accelerating phase. The universe however emerged out from an Einstein static phase. It is also shown that the static Einstein phase can be realized in the context of wormhole. The throat of the wormhole expands to a big universe after a pretty long time determined by a cosmological  term $\Lambda$ \cite{as1}. 

	For an  emergent universe  with scale factor given by eq. (\ref{e42}),  the differential equation for the Hubble parameter  is expressed as
 \begin{equation}
 \label{e15}
 \hspace{1.8 cm} \dot{H} = \alpha H - \beta H^2
 \end{equation}
 with $\alpha = \frac{ \sqrt{3}B}{2}  $ and $\beta= \frac{3(A+1)}{2}$. 
Now the scale factor  given by eq. (\ref{e42}) can be rewritten in a simple form as  \cite{eugb}
\begin{equation}
 \label{e16}
 \hspace{1.8 cm}  a(t)  = a_o \left[ \eta + e^{\alpha t} \right]^{\frac{1}{\beta}}
 \end{equation}
 denoting $a_o= \left(\frac{3K(A+1)}{2} \right)^{\frac{2}{3(A+1)}} \frac{2}{\sqrt{3} B}$, $\eta = \frac{\sqrt{3} B \sigma}{2}$. In this case  at infinitely past {\it i.e.} $t \rightarrow - \infty$, the scale factor  becomes $a \rightarrow a_o \eta^{\frac{1}{\beta}}$, thus  a static Einstein universe emerged which later evolves to a dynamical universe \cite{euorg,bas} encompassing the present observed universe.
The Hubble parameter is given by $H= \frac{\alpha e^{\alpha t}}{\beta (\eta+ e^{\alpha t})}$.

 
 The general solution of eq. (\ref{e11}) using eq. (\ref{e12}) can be obtained which accommodates emergent universe solution in a modified gravity  in vacuum which is given by
\begin{equation}
 \label{e18}
 P(t)  = - \frac{ \eta \beta^2 (\eta+2 e^{\alpha t})}{\alpha^2 (1+ \beta) e^{2 \alpha t}}  + C_1 a_o^{1+2 \beta} \int \frac{(\eta+e^{\alpha t})^{2+\frac{1}{\beta}}}{e^{2 \alpha t}}  dt +C_2
 \end{equation}
and consequently  from eq. (\ref{e10}) one obtains
\begin{equation}
 \label{e19}
 Q(t)  = - \frac{6 \alpha^2 e^{\alpha t}}{\beta^2 (\eta+ e^{\alpha t})^2} \left[ 8 \eta \beta + e^{\alpha t }\right]  + \frac{24 C_1a_o^{1+2\beta} \alpha^3 e^{\alpha t}}{\beta^3 (\eta+e^{\alpha t})^{1-\frac{1}{\beta}}}
  \end{equation}
  where $C_1$ and $C_2$ are integration constants.
 The Gauss-Bonnet combination can be expressed as a function of cosmic time $t$ which is now given by
 \begin{equation}
 \label{e20}
 G(t)  = \frac{24 \alpha^4}{\beta^4}  \; \frac{e^{3 \alpha t} ( \eta \beta + e^{\alpha t})}{(\eta+ e^{\alpha t})^4}.
 \end{equation}
 The Gauss-Bonnet terms  $G$  is a function of $t$ which is highly non-linear.  An inverse  function of $t$ in terms of $G$ namely,  $t=t(G)$ can be obtained which is given by
 \begin{equation}
 \label{e21}
  t  = \pm \left[  \frac{\ln \eta}{\alpha} - \frac{1}{\alpha} \ln \left( \left(\frac{24 \alpha^4}{\beta^4 G} \right)^{\frac{1}{4}} \mp 1)  \right) \right].
 \end{equation}

 Finally, the modified Gauss-Bonnet function is obtained which is given by
\[
 F(G) = -  \sqrt{\frac{3 G}{2}}    - \frac{2 \beta^2}{\alpha^2 (1+\beta)} \left( \frac{\alpha}{\beta} \left( \frac{24}{G} \right)^{\frac{1}{4}} -1 \right) G    
 \]
  \begin{equation}
 \label{e22}
 - \frac{4 \beta \eta^2 \sqrt{6 G}}{ \left( \frac{\alpha}{\beta} \left( \frac{24}{G} \right)^{\frac{1}{4}} -1 \right) }  - \frac{\beta^2 G}{ \alpha^2 (1+\beta) \left( \frac{\alpha}{\beta} \left( \frac{24}{G} \right)^{\frac{1}{4}} -1 \right) }  
 \end{equation}
  for $C_1=0 $ and $C_2=0$.
The interesting feature of the emergent universe solution is that the initial singularity does not arise here. The universe in course of its evolution admits observed features of the universe scenario as obtained in Ref. \cite{euorg,bas}.
 \\

 \section{. \; Emergent Universe from Mimetic $F(G)$  Gravity}
 
To obtain the mimetic $F(G)$-gravity the following parametrization of metric is assumed in the gravitational  action given by eq. (\ref{e1}) :
\cite{ca}-\cite{go6}
\begin{equation}
g_{\mu \nu} = - \bar{g}^{\rho \tau} \partial_{\rho} \phi \; \partial_{\tau} \phi \;  \bar{g}_{\mu \nu}
\end{equation}
Varying the metric one gets
\[
\delta g_{\mu \nu}= \bar{g}^{\rho \tau} \delta{g}_{\tau\omega} \bar{g}^{\omega \sigma} \partial_{\rho} \phi  \; \partial_{\sigma} \phi  \; \bar{g}_{\mu\nu} - \bar{g}^{\rho\sigma} \partial_{\rho}\phi \partial_{\sigma}\phi  \; \delta \bar{g}_{\mu\nu} 
\]
\begin{equation}
\;\;\;\;\;- 2\bar{g} ^{\rho\sigma}\partial_{\rho} \phi \; \partial_{\sigma} (\delta \phi) \;  \bar{g}_{\mu\nu}.
\end{equation}

Consequently one obtains following field equations \cite{ca,ca1,go,go1,go2,go3,go4,go5,go6}  varying the action with respect to  the redefined metric $\bar{g}_{\rho\sigma}$ instead of the standard Jordan frame metric $g_{\mu \nu}$ one obtains 
\[
R_{\mu \nu} - \frac{1}{2}  g_{\mu \nu} R + 
\]
\[
8\left[ R_{\mu \delta \nu \sigma} + R_{\delta \nu}  g_{\sigma \mu} - R_{\delta \sigma} g_{\nu \mu} - R_{\mu \nu}  g_{\sigma \delta} + R_{\mu \sigma} g_{\nu \delta}  \right]  \triangledown^{\delta} \triangledown^{\sigma} F_G  
\]
\[
+4 ( g_{\mu \sigma}  -g_{\nu \delta} ) R  \triangledown^{\delta} \triangledown^{\sigma} F_G 
+(F_G G - F(G)) g_{\mu \nu}  + \partial_{\mu}  \phi     \partial_{\nu}  \phi 
\]
\[  \left[ -R +8 \left( -    R_{\delta \sigma}   + \frac{1}{2}  g_{\delta \sigma} R  \right) 
  \triangledown^{\delta} \triangledown{\sigma} F_G            + 4(F_G G - F(G)) \right]
  \]
  \begin{equation}
\label{e20}
\; \; \; \; \; \; \; = T_{\mu \nu}+  \triangledown_{\mu} \phi \triangledown_{\nu} \phi \; T 
\end{equation}
where $F_G=\frac{dF}{dG}$ and  $ \triangledown_{\mu} f_{\nu} = \partial_{\mu} f_{\nu} - \Gamma^{\lambda}_{\mu\nu} f_{\lambda}$ with the energy momentum tensor  $T_{\mu\nu} = diagonal( \rho, -p, -p, -p)$, $\rho$  the energy density, $p$ the cosmic pressure and $\phi$ mimetic  scalar field.
Now varying the action w.r.t. $\phi$ one obtains
\[
\triangledown^{\mu} \left[ \partial_{\mu}  \phi      \left( -R - 8 \left( R_{\delta \sigma}   - \frac{1}{2}  g_{\delta \sigma}  R \right)    \triangledown^{\delta} \triangledown^{\sigma} F_G      \right)\right]
\]
\begin{equation}
\label{e21}
 +     4(F_G G - F(G))  - T = 0.
\end{equation}
From eq. (24) one obtains 
\[
g^{\mu \nu} \partial_{\mu} \phi \; \partial_{\nu} \phi = -1,
\]
where homogeneous  scalar field $\phi$ depends on  cosmic time with a constraint $\phi =t$. For a flat universe  the time-time component of the field equation  in eq. (26) is expressed as
\[
2 \dot{H} + 3 H^2 +16 H (\dot{H} + H^2) \frac{dF_G}{dt} 
\]
\begin{equation}
\label{e22}
+ 8 H^2 \frac{d^2F_G}{dt^2} - (F_GG-F(G)) = - p.
\end{equation}

On integration eq. (\ref{e21}), one obtains
\[
- R - 8 \left( R_{\delta \sigma} - \frac{R}{2}  g_{\delta \sigma}  \right)   
\partial^{\delta}   \partial^{\sigma} F_G  + 
\]
\begin{equation}
\label{e23}
   4(F_G G - F(G))  - \rho - 3 p = - \frac{C}{a^3}
\end{equation}
which further can be rewritten as
\[
2 \dot{H} + 2 H^2 + \frac{2}{3} (F_GG-F(G)) 
+
\]
\begin{equation}
\label{e24}
4 H (2 \dot{H} + 3H^2) \frac{dF_G}{dt} + 4 H^2 \frac{d^2F_G}{dt^2} = - \frac{\rho}{6} +  \frac{p}{2} -\frac{C}{a^3}
\end{equation}
here $C$ is an integration constant.
Using eqs. (\ref{e22}) and (\ref{e24}) a final equation is obtained which is
\[
\dot{H} + 2 H^2 + \frac{2}{3} (F_G \; G-F(G)) 
+
\]
\begin{equation}
\label{e25}
4 H (2 \dot{H} -  H^2) \frac{dF_G}{dt} + 4 H^2 \frac{d^2F_G}{dt^2} = - \frac{1}{2}  (\rho +p) -\frac{C}{a^3}.
\end{equation}
Here we denote $g(t) = \frac{dF_G}{dt}$ and the above equation can be written as
\begin{equation}
\label{e26}
4 H^2 \frac{dg(t)}{dt}     + 4 H (2 \dot{H} - H^2) g(t)    = - \dot{H} - \frac{(\rho +p)}{2}   -\frac{C}{a^3}.
\end{equation}
To obtain an emergent universe model discussed in sec.(II), we use the eq. (\ref{e15}), it will generate the modified gravity.  The general solution now can be obtained from eq.(\ref{e26}) corresponding to a functional form given by
\begin{equation}
\label{e27}
g(t)= -  \frac{C_1}{H^2}  a(t) 
\end{equation}
where $C_1$ is an integration constant. In this section, we reconstruct  the modified Gauss-Bonnet gravity to realize the emergent universe scenario obtained by Mukherjee {\it et. al} \cite{euorg} in Einstein's General theory of relativity with non-linear equation of state describing the matter. 
 Now emergent universe model is obtained when the right hand side of eq. (\ref{e26}) in the absence of dark radiation ($C=0$) is replaced by
 \begin{equation}
 \dot{H} + \frac{1}{2}  (\rho +p) = \frac{1}{2} H^2
 \end{equation}
 which determines the modified Gauss-Bonnet terms. The modified GB terms will be determined in the next section.
 
The Hubble parameter for emergent universe model is 
\[
H = \frac{H_0 e^{\alpha t}}{\eta+ e^{\alpha t}},   \; \; \; H_0= \frac{\alpha}{\beta}.
\]
Thus the general solution in terms of $t$ is given by
\begin{equation}
g(t) = \frac{C_1 a_o \beta^2}{\alpha^2 e^{2\alpha t} } \left( \eta +    e^{\alpha t} \right)^{2+\frac{2}{\beta}}.
\end{equation}  
The modified Gauss-Bonnet gravity is then determined in terms of $t$ as
\begin{equation}
F_G (t) = \int g(t) dt.
\end{equation} 
It can be integrated for different values of  $\beta $. 

$\bullet$ For $\beta = 1$ one obtains
\[
F_G (t)=  g_o \left(  \frac{e^{2 \alpha t}}{2}    + 4 \eta  e^{\alpha t}  +6 \eta^2  \alpha \; t - 4 \eta^3 
e^{- \alpha t}  -  \frac{\eta^4 e^{- 2 \alpha t}}{2}  \right) 
\]
\begin{equation}
\hspace{4.8 cm} +F_o
\end{equation}
where $g_o =\frac{C_1 a_o\beta^2}{\alpha^3  }  $ and $F_o$ is a constant.

$\bullet $ For $\beta =2$
\begin{equation}
F_G (t) = g_o   \left(  e^{\alpha t}    + 3  \eta \alpha \; t - 3 \eta^2 
e^{- \alpha  \; t} -  \frac{\eta^3}{2} e^{- 2 \alpha \; t} \right) +F_1
\end{equation}
where $F_1$ is an integration constant. It is noted that $\beta$ is playing here an important role in determining the modified Gauss-Bonnet gravity to accommodate emergent universe solution. 
Using  the GB terms with the scale factor for emergent universe, $t$ can be replaced by $G$ as
\begin{equation}
t= \frac{1}{\alpha} \frac{\eta}{\left( \frac{G_o}{G} \right)^{1/4} -1}
\end{equation}
where $G_o= 24 H_0^4 (1+\beta \eta)$.
The  functional form of modified Gauss-Bonnet gravity can be determined from \\

(i) when $\beta=1$
\[
F_G (G) = g_o \left( \frac{\eta^2}{2   \left[   \left(  \frac{G_o}{G}  \right)^{\frac{1}{4}} -1  \right]^{2}} + \frac{4 \eta^2}{\left(  \frac{G_o}{G}  \right)^{\frac{1}{4}} -1} \right)
\]
\[
\; \; \; -6 \eta^2 g_o\ln  \left[ \left(  \frac{G_o}{G}  \right)^{\frac{1}{4}}  -1\right] 
-  4 g_o \eta^2 \left[ \left(  \frac{G_o}{G}  \right)^{\frac{1}{4}} -1  \right] 
\]
\begin{equation}
\label{e40}
- \frac{g_o \eta^4}{2}  \left[ \left(  \frac{G_o}{G}  \right)^{\frac{1}{4}} -1  \right]^{2}    +f_o
\end{equation}

(ii) when  $\beta=2$
\[
F_G (G) = g_o \left( \frac{\eta}{   \left(  \frac{G_o}{G}  \right)^{\frac{1}{4}} -1  } -3 \eta \ln \left[
\left(  \frac{G_o}{G}  \right)^{\frac{1}{4}} -1  \right]
\right)
\]
\begin{equation}
\label{e40}
-  g_o \eta  \left[ \left(  \frac{G_o}{G}  \right)^{\frac{1}{4}} -1  \right]     -    \frac{g_o \eta}{2}  \left[ \left(  \frac{G_o}{G}  \right)^{\frac{1}{4}} -1  \right]^2 +f_1
\end{equation}
where $f_o$ and $f_1$ are constants. The above modified gravitational actions  accommodates EU scenario fairly well.
For $G >>  G_o$  one gets

$\bullet$  $ F(G) = \lambda_1 +  \left(   f_o -10 g_o \eta^2 - \frac{1}{2} g_o \eta^4 - \frac{ 3 \eta^2 g_o \ln G_o }{2} \right) G + \frac{3 \eta^2 g_o}{2} \;G \; \ln G - \frac{13 g_o \eta^2 G_o}{2}  \; \ln G + 8 \eta^2 g_o G_o^{\frac{3}{4}} \; G^{\frac{1}{4}} -(2 \eta^2 +11) \;  g_o \eta^2 \sqrt{G_o \; G} + 4 g_o \eta^2 (\eta^2-1) G_o^{\frac{1}{4}} \; G^{\frac{3}{4}} ; $

$\bullet$
$ F(G) = \left( f_1 - \frac{3g_o \eta}{2} - \frac{3g_o \eta}{4} \ln G_o \right) \; G + \frac{3 g_o \eta}{4} G \; \ln G - g_o \;  \eta \; G_o \;  \ln G + 4 g_o \;  \eta \; G_o^{\frac{3}{4}}  \; G^{\frac{1}{4}} - 3 g_o \; \eta \; \sqrt{G_o}  \; \sqrt{G} + \frac{4 g_o \; \eta \; G_o^{\frac{1}{4}}}{3} \; G^{\frac{3}{4}} + \lambda_2 $ \\

where $\lambda_1$ and  $\lambda_2$ are integration constants for $\beta =1$ and $2$ respectively.

\vspace{0.8 cm}

\section{. \; Discussion}
  
In this paper modified gravity with Gauss-Bonnet terms is reconstructed which  accommodates flat emergent universe model. Earlier a number of cosmological  models with various evolution having a bouncing solution are explored with a reconstructed mimetic gravity \cite{asc,bam9,kb}, but the present work is different. It is known that the  Gauss-Bonnet terms in four dimensions is equivalent to a Euler number which does not contribute in the dynamics of a matter dominated universe. However, in the presence of a dilaton field coupled to Gauss-Bonnet combination is capable of producing the evolutionary features of the observed universe.  It is also shown that an Emergent universe scenario can be realized in this framework which describes early universe and late universe fairly well \cite{eugb}. The non-linear functional form of  the Gauss-Bonnet gravity is obtained which accommodates the emergent universe scenario with all the features of the observed universe.  We note that in the absence of a dilaton field a non-linear combination of GB terms accommodates EU solution. We reconstructed the modified gravity action with $F(G)$ which is relevant in the early and also  in the late universe. The non-linear EoS parameters $A$ and $B$ of GR is replaced here by $\alpha$ and $\beta$.  It is noted that a number of modified gravity can be predicted with different values of $\beta$-parameter  which admits EU model here. It is found that for a given $\alpha$,  the Hubble parameter decreases with an increase in   $\beta$ at the present epoch. Thus Gauss-Bonnet terms play an important role in understanding the observed universe which is described by  dark matter and dark energy  in Einstein General theory of relativity.  The parameter $\eta$ is also playing a significant  role which analogously compared with the  non-linear EoS parameter  $A$  in GR for obtaining EU scenario. Thus  a rich structure of GR modified with  non-linear combination of Gauss-Bonnet terms in the Einstein-Hilbert action is explored here which  plays an important role in cosmology.

\section{Acknowledgements:}

BCP would like to acknowledge support for a visit to the University of Zululand, South Africa and University of Kwa-Zulu Natal, South Africa where the work has been initiated.
BCP would like to thank IUCAA Pune, India for warm hospitality and DST-SERB for supporting  project (EMR/2016/005734).


\end{document}